\magnification=\magstep1
\hsize=15.5 truecm
\vsize=21.0 truecm

\leftskip=0.5 truecm
\topskip=3 truecm
\splittopskip=0 truecm
\parskip=5 pt plus 1 pt
\baselineskip=16.5 pt

\centerline {\bf SUPERSINGULAR SCATTERING}
\bigskip
\bigskip
\bigskip
\bigskip
\centerline {T. Dolinszky}
    \bigskip
\centerline {KFKI-RMKI, H-1525 Budapest 114, P.O.B. 49, Hungary}
\bigskip

\bigskip
\bigskip
\bigskip

\bigskip 
\parindent 0pt                                                       
{\bf Abstract:}
In 'supersingular' scattering the potential $g^2U_A(r)$ involves a             variable nonlinear parameter $A$ upon the increase of which the potential      also increases beyond all limits everywhere off the origin and                 develops a uniquely high level of singularity in the origin. The               problem of singular scattering is shown here to be solvable                    by iteration in terms of a smooth
version of the semiclassical approach to quantum mechanics. Smoothness
is achieved by working with a pair of centrifugal strengths                    
within each channel. In both of the exponential and trigonometric regions,     integral equations are  set up the solutions  of which when matched            smoothly may recover the exact scattering wave function. The conditions        for convergence of the  iterations involved are derived for both fixed         and increasing parameters. In getting regular scattering solutions, the        proposed procedure is, in fact, supplementary  to the Born series by           widening its scope and extending applicability  from nonsingular               to singular potentials and from fixed to asymptotically  increasing,            linear and nonlinear, dynamical  parameters.
\bigskip
\bigskip
r: 24.10 Fr... 03.65 Nk
                       
\parindent 28pt  

\vfill\eject

\topskip=1 truecm
\vsize=21.0 truecm
\bigskip
\centerline {\bf I. Introduction}
          We are going to consider scattering by singular potentials
$g^2U_A(r)$ at fixed as well as asymptotical values  of the
linear and nonlinear parameters $g^2$ and $A$, respectively.                   
The pioneering  work in treating singular scattering by exact means            has been done  by Calogero$^1$ in terms of the phase approach to               quantum particle dynamics.   
In particular, he calculated the high energy limit of scattering               by inverse power potentials$^2$. A further correct approach is due             to Froemann and Thylwe$^3$, who analytically proved the exactness of the first order WKB (Wentzel-Kramers-Brillouin) approximation for the case of pure       inverse  power potentials in the short wavelength limit. Esposito$^4$ worked   
in singular scattering by the wave function polydromy method to get a recursion formula between solutions with potentials of different inverse                powers in the radial distance. Notice all the above approaches relied on       nonlinear first order or linear  second order differential equations           solvable analytically, in terms of special  functions, for certain             simple potential shapes.          

          Nevertheless, it is the method of integral equations that is         the first rank candidate for solving two body scattering problems              involving a $general$ shape and stage of the potential singularity.
The pertinent  Volterra type  equations can be, perhaps, best  classified    
in terms of the reference scattering  problems implied , as follows.

          The most evident reference basis seems to be the unperturbed     
case $g^2=0$ taken at the physical  energy $k^2$  and the physical             
orbital angular momentum $l$. The relevant  integral equations                 set up for the regular  wave function $u^+(r)$ work                            in the case of  nonsingular potentials,
exclusively (Born series). It is true that  for finding irregular              solutions $u^-(r)$, there are analogous integral equations available,          for cases of nonsingular and singular potentials alike,                        as shown by Newton$^5$.
However, all the above mentioned procedures can only be employed  at
fixed sets of the scattering  parameters. Namely, the
residual potential contained is there invariably given by the expression       $\Delta (r)= g^2U_A(r)$ , whence the iterated series blows up               
  term-by-term in the limit $g^2\to\infty$ (strong coupling). The same is
true of  the supersingular limit. Namely,  the form factor  $U_A(r)$,          besides  being singular for $r\to 0$ at
fixed parameters, increases, by definition,  beyond                            all limits at  each fixed  point off the
origin for  $A\to \infty$. One has therefore to seek alternative approaches    outside of  the Born series for solving  these asymptotical cases.

          An interesting choice of  reference basis, as proposed by            Newton$^5$,  offers itself in the exceptional  cases of  the                   solution being known in closed form  at zero energy  and zero orbital      
angular momentum. The residual potential is then                              
$k^2-l(l+1)/{r^2}$. It is worthwhile to note that the iteration                of the integral equation converges exclusively for singular potentials.        
Moreover, convergence obviously  holds in both limits  $g^2\to\infty$
and $A\to\infty$.

          Quite wide is the scope  of applicability of the third type          
of reference problems  which  we are going to consider in detail.              It is, in fact, the semiclassical approximation.                               The Langer version of this method, see e.g. Newton's monograph$^5$,            is specified by the reference centrifugal strength  $(l+{1\over 2})^2$.        Owing to this very choice, the model wave function  reproduces the             exact regular solution of quantum   mechanics (QM)                             near the origin in the  case of nonsingular potentials. As to singular         potentials, the small distance QM  wave function and its  WKB approximation    do coincide, even independently  of the selection  of the reference            centrifugal strength. The problem of a possible convergent iteration           and that of asymptotical values taken  by the potential parameters             had remained thereby still  open. Recently, Dolinszky$^6$ proposed a           smooth version of the semiclassical approach which is free of the              highly inconvenient  turning point  singularity  inherent                      in both the Langer's and the standard WKB approximation. The new               procedure  has been applied for developing convergent series expansion          of the wave function describing scattering by singular potentials  at fixed$^7$ as well as increasing$^8$  linear dynamical  parameters.
                                                                                           In the present paper an improved  and generalized version of       that approach will be proposed, which also includes the asymptotical case       of $increasing \ \  nonlinear$  parameters present  in the Schroedinger       equation.       
\bigskip
\centerline {\bf II. Physical and reference problem}   
          The physical problem to be solved first is two-body                  scattering by a central singular potential say,  $g^2U_A(r)$                   at the energy $k^2$ in the channel of index $l$, with $A$                      representing a nonlinear parameter. The potential should                       fulfill  the following requirements:
  $$ \eqalign{ & [r^4U_A(r)]\to\infty,[r \to 0]; \ \  U_A(r)>0,                    U_A'(r)<0,[r\geq 0]; \cr & [r^3 U_A(r)] \to 0,[r\to\infty]; \ \                     U_A(r)\to\infty,[r\geq 0,A\to\infty]. \cr } \eqno(2.1) $$
In order to construct a suitable reference system for treating the 
problem, we introduce a triad  $(\lambda_\epsilon^2,\lambda_\tau^2, 
\lambda^2)$ of auxiliary centrifugal strengths, subject to the restrictions 
  $$ \lambda^2(l) = 
{1\over 2}[\lambda_{\epsilon}^2(l)+\lambda_{\tau}^2(l)],\ \ \                  
\lambda_{\epsilon}^2(l)>\lambda_{\tau}^2(l).\eqno(2.2) $$
The constants $\lambda _\epsilon^2$ and $\lambda _\tau^2$ are                  for the present freely chosen. The concept 'matching distance'                 is defined  as the positive root of the 'master equation'                    
  $$ k^2R^2-g^2R^2U_A(R)-\lambda^2=0; \ \ R=R(k^2,g^2,\lambda^2;A)>0.            \eqno(2.3) $$                                                                  The radial distance $r$ will be in general substituted for by the              dimensionless radial coordinate $t$. We put 
  $$ t={r\over R},\ \  [r \geq 0]. \eqno(2.4) $$ 
Different regions of the space are  distinguished as follows: 
  $$ \eqalign{ t&<1, \ \ \rm{region} \ \ \epsilon, \ \                          (\rm{exponential \ \  region});\cr 
t&>1, \ \  {\rm region} \ \ \tau,  \ \ ({\rm trigonometric \ \  region}); \cr 
t&=1, \ \ ({\rm matching \ \  point}).\cr} \eqno(2.5) $$
The scattering process  is governed by the radial Schroedinger equation
  $$ \{{{\rm d}^2 \over {\rm d}t^2}+k^2R^2 -g^2R^2U_A(Rt)-{l(l+1)\over t^2}\}      u^{\pm}(t)=0. \eqno(2.6) $$
A regular-irregular pair  $u^{\pm}(t)$ of its  solutions behaves               near the origin as
  $$ u^{\pm}(t)\to [g^2 R^2 U_A(Rt)]^{-{1\over 4}}\exp[\pm gR
\int_1^t {\rm d}t'U_A(Rt')^{1\over 2}],\ \ \   [t\to 0].\eqno(2.7) $$

    The reference problem will be a special  smooth version of the 
zero-order  semi-classical approximation. The entire argument hinges upon a 
pair of wavenumber function squares for the exponential and trigonometric
regions, respectively, as follows:
  $$ \eqalign {K^2(t) &=K^2_\epsilon (t) = -k^2+g^2U_A(Rt)+
{\lambda_\epsilon ^2 \over R^2t^2}, \ \ [t<1]; \cr & =  K^2_\tau (t)            = k^2-g^2U_A(Rt)-
{\lambda_{\tau}^2 \over R^2t^2},\ \ \ [t>1].\cr} \eqno (2.8) $$
These definitions imply a number of properties concerning  behavior         
around the matching point. In particular, one extracts                        from the master equation $(2.3)$ that       
  $$ \bigl[K_{\epsilon}^2(t)\bigr]_{t=1}=\bigl[K_{\tau}^2(t)\bigr]_{t=1};
\ \ \  [K^2(t)]_{t=1}= { \lambda_\epsilon^2-\lambda_{\tau}^2 \over 2R^2}.
\eqno (2.9) $$
Overall properties also follow from Eq.(2.8). Indeed,
  $$ \eqalign {K^2(t) &>0,\ \ \ \min\{K^2(t)\}= \bigl[K^2(t)\bigr]_{t=1},      \ \ \ [0\leq t]; \cr K^2_\epsilon(t_1) &>K^2_\epsilon(t_2) \ \                 {\rm if} \ \ t_1 <t_2; \ \ \  
 K^2_\tau(t_3)< K^2_\tau(t_4) \ \ {\rm if} \ \ t_3<t_4.\cr} \eqno (2.10) $$ 
These statements issue from  the restrictions (2.1) imposed on $U_A(r)$.       
Small and large distance behavior can also  be thence extracted such as
  $$ K^2_\epsilon(t)\to \infty \ \ {\rm if} \ \ t\to 0; \ \   K_\tau^2(t) \to
k^2 \ \ {\rm if} \ \ t \to \infty. \eqno(2.11) $$
For the derivatives of $K^2(t)$ we shall apply the notation:
   $$ D_\gamma^{(s)}(t)= {1\over K^2_{\gamma}(t)} {{\rm d}^s                   K^2_\gamma(t)\over
{\rm d}t^s},\ \ \ [s=1,2;\gamma= \epsilon, \tau]. \eqno (2.12) $$ 
Hence one obtains by Eqs.(2.1)-(2.2) and (2.8) 
  $$ \bigl[D^{(1)}_\tau(t)-{D^{(1)}_\epsilon (t)}\bigr]_{t=1}=
2^4\{2\lambda^2-g^2R^3 U_A'(R)\}>0. \eqno(2.13) $$
          A regular-irregular pair of reference wave functions                 $w^\pm_\gamma(t)$ is defined  as
  $$ w^{\pm}_{\epsilon}(t)=\eta_{\epsilon}(t) \exp  
[\pm \omega _\epsilon (1,t)],  \ \ \ [t<1]; \eqno(2.14) $$  
  $$ w^{\pm}_{\tau}(t)=\eta_\tau (t)[C^{\pm}\cos \omega_{\tau}(1,t))             + S^{\pm} \sin \omega _{\tau}(1,t) ],\ \ [t>1], \eqno (2.15) $$
where we introduced the 'amplitude function' and the 'phase function' such as 
  $$ \eta _{\gamma}(t)\equiv \left({k^2\over K^2_\gamma(t)}\right)^{1\over 4}, \ \ [\gamma=\epsilon,\tau]; \eqno(2.16) $$
  $$ \omega_{\gamma}(t_1,t_2) \equiv R \int_{t_1}^{t_2}                        {\rm d}t' |K_{\gamma}(t')|,        
\ \ \ [\gamma =\epsilon, \tau]. \eqno(2.17) $$
As to the constants $C^\pm$ and $S^\pm$ , these parameters  can be,         
for the moment, freely chosen. Out of them, $C^+$ and $S^+$ will               be specified later  by smoothness requirements. The choice $C^-$               and $S^-$ will, in turn, remain once for all free but the only restriction
  $$ C^+S^-S^+C^- \not= 0, \eqno(2.18) $$
which warrants  independence of the functions $w_\tau^+(t)$ and                $w_\tau^-(t))$. Yet, there exists a sophisticated definition of 
the constants $C^-$ and $S^-$ in terms of $C^+$ and $S^+$, namely the one      implied in the identity 
  $$ w_\gamma^-(t) \equiv w_\gamma^+(t) \{1- \int_1^t {{\rm d}t' 
\over w_\gamma^+(t')^2 } \}, \ \ \ [\gamma =\epsilon, \tau].                   \eqno(2.19) $$
This relationship is automatically satisfied in the region $\epsilon$          by  the definition (2.14). As regards  the point $t=1 \pm 0$, Eq.(2.19)        guarantees there  smooth matching of 
$w_\tau^-(t)$ to $w_\epsilon^-(t)$ whenever $w_\tau^+(t)$ matches 
there $w_\epsilon^+(t)$ smoothly. The particular choice (2.19) is 
quite irrelevant  from the viewpoint of the present  argument.                 Nevertheless, it justifies the use of the same superscripts                     $(^-)$ over ${w_\tau}^-(t)$ and  ${w_{\epsilon}}^-(t)$.

          The reference wave functions $w_\gamma^\pm(t), \ \ [\gamma=\epsilon, \tau]$ , of Eqs.(2.14)-(2.15)             
solve the pair $(\epsilon,\tau)$ of differential equations
  $$ \{ {{\rm d}^2 \over {\rm d}t^2} +k^2R^2                                   - W^{\pm}_{\gamma}(t) -{l(l+1)\over t^2} \}w^{\pm}_\gamma(t)=0,\ \ \ \            [\gamma=\epsilon,\tau]. \eqno(2.20) $$
Notice the differential equation (2.20 is common for $w^+_{\gamma}(t)$ and 
$w^-_{ \gamma}(t)$, in the exponential and  trigonometric                      regions alike. Namely, calculation yields for the reference potential  
  $$ W^{\pm}_\gamma(t) \equiv W_{\gamma}(t),\ \ \ [\gamma=\epsilon, \tau].
\eqno (2.21) $$                                                                with the notation                        
  $$ W_\gamma(t)=g^2R^2U_A(Rt) + \Delta _\gamma(t),  \ \ \                            [\gamma =\epsilon,\tau].\eqno(2.22) $$
The expressions  of the residual potential $\Delta_\gamma(t)$ introduced       here are extracted from Eqs.(2.14)-(2.15) and (2.20)  as
  $$ \Delta_\epsilon(t) \equiv -{5 \over 16} [D^{(1)}_{\epsilon}(t)]^2
+ {1\over 4} D^{(2)}_{\epsilon}(t)-{\lambda ^2_{\epsilon}(t) -{l(l+1)}              \over{t^2}}, \ \ \ [t<1], \eqno(2.23) $$
  $$ \Delta_\tau (t) \equiv -{5 \over 16}[D^{(1)}_{\tau}(t)]^2                 +{1\over 4} D_{\tau}^{(2)}(t)-{\lambda ^2_\tau-
l(l+1)\over {t^2}}, \ \ \ \ [t>1]. \eqno (2.24) $$
Recall that  $K^2(t)$ of Eq.(2.8) is by Eqs. (2.9) continuous at $t=1$         but, due to (2.12)-(2.13), not smooth. Therefore, the residual potential        develops at the matching point a jump. Indeed,
  $$ \Delta_{\tau}(t+0)-\Delta _{\epsilon}(t-0)\not= 0. \eqno(2.25) $$        
For simplicity, see expression (2.24), we shall work hence forward              with the centrifugal strengths
  $$ \lambda_\epsilon^2=(l+{1\over 2})^2, \ \ \  \lambda_\tau^2=l(l+1).        \eqno(2.26) $$
Observe that this choice is compatible  with the inequality contained in 
the postulates (2.2).       
 \bigskip
  \centerline  {\bf III. A pair of convergent expansions}
          A comparison of the Schroedinger equations  (2.6), set up for        the exact wave function $u^+(r)$, and Eq.(2.20), solved by the semiclassical   wave function $w^+(r)$, suggests  construction of a  pair of                   integral equations,
  $$ v_{\epsilon}^+(t)=w_{\epsilon}^+(t)+\int_0^t {\rm d}t'\Delta               _{\epsilon}(t') G_{\epsilon}^+(t,t')v_{\epsilon}^+(t'), \ \ \ [t<1],          \eqno(3.1) $$
  $$ v_{\tau}^+(t)=w_{\tau}^+(t)+\int_1^t{\rm d}t'                             \Delta_{\tau}(t') G_{\tau}^+(t,t') v_{\tau}^+(t'), \ \ \ [t>1].\eqno(3.2)$$
The solutions $v_{\epsilon}^+(t)$ and $v_{\tau}^+(t)$, if exist, are           solutions of the exact Schroedinger equation within the respective regions.    In particular, $v_{\epsilon}^+(t)$ is uniquely defined  by Eq.(3.1)            and furnishes a  regular solution of the differential equation (2.6)           in the exponential region. The solution $v_{\tau}^+(t)$ of  Eq.(3.2),           in turn, while solving the Schroedinger equation  in the                       trigonometric region, still involves two free constants, $C^+$ and $S^+$,       as noticed following Eq.(2.15). Smoothness  requirement for the               overall regular solution $v^+(t)$ at  $t=1$  is just sufficient to             unequivocally specify these coefficients. As to the notation in (3.1)-(3.2),   the residual potentials $\Delta_\gamma(t)$  have been defined by (2.22)-(2.24). The resolvents involved in the integral equations are formally given as
  $$ G_{\gamma}^+(t,t')= {1\over d_{\gamma}^+} [w_{\gamma}^+(t)w_{\gamma}^-(t')
-w_{\gamma}^-(t)w_{\gamma}^+(t')],\ \ \ [\gamma=\epsilon,\tau],  \eqno(3.3) $$
where the Wronskians contained are by Eq.(2.21) constant and read in general
  $$ d_{\gamma}^+= {\rm W}_{\gamma}\{w_{\gamma}^+(t); w_{\gamma}^-(t) \} =
{\rm const.}, \ \ [\gamma=\epsilon,\tau]. \eqno(3.4a) $$                        In particular, one obtains after  some calculations
  $$  d^+_\epsilon =-2kR, \ \ \ d^+_\tau=  kR(C^+S^--C^-S^+). \eqno(3.4b) $$   
The general expressions (3.3) can be recast in terms                           of the local wave numbers as
  $$ G^+_{\epsilon}(t,t')= 2 {\sinh [\omega_{\epsilon}(t,t')] \over 
R[K^2_{\epsilon}(t)K^2_{\epsilon}(t')]^{1\over 4}},\ \ \ \ 
[0 \leq t' \leq t <1], \eqno(3.5) $$
  $$ G^+_{\tau}(t,t')={\sin [ \omega _{\tau}(t,t')]
\over R[K^2_{\tau}(t)K^2_{\tau}(t')]^{1\over 4} },                                 \ \ \ \ [1 \leq t'\leq t].\eqno(3.6) $$
Irrelevance of any special  choice of the basis $w^{\pm}_{\tau}(t)$            for inclusion in the formula (3.3) manifests  itself in the absence            of  the constants $C^{\pm},S^{\pm}$ from the last  expression.
                    
          The solution of the integral equations (3.1)-(3.2)                   rests upon the recursion schemes 
  $$ w^+_{\epsilon n}(t) \equiv \int_0^t {\rm d} t' \Delta _{\epsilon}(t') 
G^+_{\epsilon}(t,t') w^+_{\epsilon n-1}(t'), \ \ [n=1,2,3..]; \ \ \
w^+_{\epsilon 0}(t) \equiv w^+_{\epsilon}(t),\eqno (3.7) $$
  $$ w^+_{\tau m}(t) \equiv \int_1^t {\rm d}t'\Delta_{\tau}(t')                G^+_{\tau}(t,t') 
w^+_{\tau m-1}(t'), \ \ [m=1,2..]; \ \   w^+_{\tau 0}(t)                    
 \equiv w^+_{\tau}(t).\eqno(3.8) $$                                              A necessary condition for getting the solution sought for                      by iteration  is convergence of each of the infinite  series
  $$ v^+_{\epsilon}(t)= \sum_{n=0}^\infty w^+_{\epsilon n}(t), \ \ \ [t<1];
\eqno (3.9a) $$  
  $$ v^+_{\tau}(t)= \sum_{m=0}^\infty w^+_{\tau m}(t), \ \ \                   [t>1].\eqno(3.9b) $$
The overall solution $v^+(t)=\{v^+_\epsilon(t);v^+_\tau(t)\}$  should be             throughout smooth. This requirement is realized off the matching               point spontaneously. At the matching point, it is the zero order term of
the series (3.9b) that exclusively contributes to both  the solution           $v^+_\tau(t)$ and  its first derivative. If the series (3.9a) is in             the region $\epsilon$ convergent, the smoothness postulate                    for $t=1$  can  be recast in a simple form as
  $$ [w^+_\tau(t)]_{t=1}= [v^+_\epsilon(t)]_{t=1}; \ \ 
     [w^+_\tau(t)']_{t=1}= [v^+_\epsilon(t)']_{t=1}.\eqno(3.10) $$
These  conditions simultaneously fix the trigonometric constants $C^+$
and $S^+$ thus completing the solution in the region $\tau$.                           
          The convergence proof for the above series  becomes                   more transparent  by using the following  notation in both regions
 $\gamma=\epsilon,\tau$
  $$ q^+_{\gamma s}(t) \equiv { w^+_{\gamma s}(t) \over                        w^+_{\gamma 0}(t)}; \ \ \ q_{\gamma 0}^+(t)=1, \ \ \eqno(3.11) $$              
  $$ p_{\gamma}(t) \equiv {\Delta_{\gamma}(t) \over  RK_{\gamma}(t)},           \eqno(3.12) $$
  $$ P_\gamma (t_1,t_2) \equiv \int _{t_1}^{t_2} {\rm d}t'|p_\gamma (t')|.
\eqno(3.13) $$                                                               . In the exponential region, the formula (3.7) can thus            
be rewritten as             
  $$ q^+_{\epsilon n}(t)= \int_0^t{\rm d}t'p_\epsilon(t')                      \{1- \exp [2\omega_{\epsilon}(t,t')]\} q^+_{\epsilon n-1}(t'). 
\eqno(3.14) $$
Hence we get by Eq.(2.17) the inequality
  $$ |q^+_{\epsilon n }(t)| \leq  \int_0^t {\rm d} t'|p_{\epsilon}(t')      q^+_{\epsilon n-1}(t')|, \ \ \ [t <1]. \eqno(3.15) $$
                  Iteration yields then by Eq.(3.13) 
  $$ |q_{\epsilon n}(t)|\leq  {1\over n!}[ P_{\epsilon}(0,t)]^n, \ \ \ [t<1],
\eqno (3.16) $$                                                                
whence  one extracts  by analysis
  $$ \sum_{n=0}^\infty |w_{\epsilon n}^+(t)|< |w_{\epsilon}(t)|                
\exp [P_{\epsilon} ( 0,t)], \ \ \  [t<1]. \eqno (3.17) $$
The series $v_{\epsilon}^+(t)$ of Eq.(3.9a) is thus absolutely                  convergent if and only if the integral $P_{\epsilon}(0,t)$                     exists and is bounded  in $t=(0,1)$.
As to the trigonometric region, the resolvent formula (3.6) combines with                 the notation (3.12) to an equivalent form of the recursion relationship (3.8). Accordingly, we get
  $$ w_{\tau m}^+(t) =\int_1^t {\rm d}t' p_{\tau}(t') \sin \omega_\tau (t,t')
{K_{\tau}^{1\over 2}(t')\over K_{\tau}^{1\over 2}(t)}                          w_{\tau m-1}^+(t'), \ \ \ [t>1]. \eqno (3.18) $$
The monotonicity relationship (2.10) implies then the inequality
  $$ |w_{\tau m}^+(t)| < \int_1^t {\rm d} t'|p_{\tau }(t')|                      |w_{\tau m-1}^+(t')|, \ \ [t>1, m=1,2,3...]. \eqno(3.19) $$
Recall now the definition (2.15) and conclude by the zero order, $m=0$,        identity in the relationships (3.8) that
  $$ |w_{\tau 0}^+(t)| < (kR)^{1\over 2} (|C^+|+|S^+|), \  \  [t \geq 1] .\eqno(3.20) $$ 
Iteration furnishes  then  by the inequalities  (3.19)-(3.20)
  $$ |w_{\tau m}^+(t)| < (kR)^{1\over 2}(|C^+|+|S^+|) {1\over m!}               [P_{\tau}(1,t)]^m, 
\ \ [t>1;m=1,2,3...]. \eqno(3.21) $$
Summation over $m$ yields thus  
  $$ \sum_{m=0}^{\infty} |w_{\tau m}^+(t)|< (kR)^{1\over 2}(|C^+|+|S^+|)       
                 \exp P_\tau (1,t) , \ \ [t>1]. \eqno(3.22) $$
So the series $v_{\tau}^+(t)$ of the definition (3.9b) is absolutely           convergent whenever  the integral $P_{\tau}(1,t)$ of Eq.(3.13) does exist.

          Suppose the regional existence conditions
  $$  P_\epsilon(0,t)<\infty,\ \ [t<1], \ \ \  P_\tau(1,t)<\infty, \ \ [t>1],
\eqno(3.23) $$
and, in addition, the smoothness postulate (3.10) are satisfied. If so,    
then solutions of the integral equations (3.1)-(3.2) together  recover         the regular solution of the differential equation (2.6). In particular,
  $$  u^+(t)=v^+_\epsilon(t), \ \ [t<1]; \ \ \ u^+(t)=v^+_\tau(t),
 \ \ [t>1]. \eqno(3.24) $$  
Virtually, one always works , instead of the infinite  expansions
(3.9a)-(3.9b), with cut-off series such as 
  $$  v_\epsilon^{+(N)}(t)= \sum_{n=0}^N w_{\epsilon n}^+(t), \ \ [t<1];
\eqno(3.25) $$
  $$  v_\tau^{+(NM)}(t)=\sum _{m=0}^M w_{\tau m}^{+(N)}(t), \ \ [t>1].
\eqno(3.26) $$ 
Smooth matching of these functions at $t=1$ in terms of the relevant           analog of  Eqs. (3.10)  fixes the trigonometric
coefficients $C^+$ and $S^+$ in terms of the cut-off 'length' $N$ in the       
exponential region but independently of its pair $M$ in the trigonometric      region.  These constants therefore should  carry the superscripts $(N)$        only. The postulates (3.10), rewritten  for  the cut-off approach,  read 
  $$  2^{3\over 4} (kR)^{1\over2} C^{+(N)}= [v^{+(N)}_\epsilon(t)]_{t=1},
\eqno(3.27) $$
  $$  -2^{-{5\over 4}}                                                         
[D_\tau^{(1)}(t)]_{t=1}(kR)^{1\over 2}C^{+(N)}
+ 2^{-{3\over 4}} S^{+(N)}= [v_\epsilon^{+(N)}(t)']_{t=1}. \eqno(3.28) $$
Notice the cut-off approach must not be used unless  the conditions of         convergence 
 are fulfilled in both regions $\epsilon$ and $\tau$.
\bigskip
\centerline {\bf IV. The supersingularity  limit }
          The term 'supersingularity' implies in our terminology  a singular
potential which is subject to the requirements (2.1). This involves, among     others, that it contains a nonlinear parameter $A$ which may increase          beyond  all limits. The relevant asymptotical form of                           the master equation  (2.3) reads 
  $$  k^2-{g^2\over r_0^2} U_A(R_A)- {\lambda^2\over R_A^2}\to 0,             \ \ [A\to \infty].\eqno(4.1) $$                                                 Bounded values such as $R_A\to 0$ and $R_A\to {\rm const.}$ are by             the properties (2.1) obviously excluded from the large-$A$  solutions          of the Eq.(4.1). One is thus left with the only possibility that               $R_A\to\infty$ for $A\to\infty$. The master equation itself reduces therefore  in the supersingularity limit  to
  $$ U_A(R_A)\to {k^2r_0^2\over g^2}, \ \ [A\to\infty]. \eqno(4.2) $$     
It is perhaps worth recalling  that $U_A(r)\to 0 \ \ {\rm if} \ \                A= {\rm fixed},r\to \infty$ while  $U_A(r)\to \infty \ \ {\rm if} \ \ 
r={\rm fixed},A\to\infty$, as implied in the set of postulates (2.1). In Eq.   (4.2), in turn, the singularity parameter $A$ and the matching radius  $R_A$
vary simultaneously  and  both increase to infinity so as to keep              the left hand side of the equation constant.

                Four classes of strongly singular potentials will                be introduced   with each potential being  specified by a variable           core parameter $A$ and a fixed tail parameter $B$. The dimensionless           form factor is in each case a product of an, for $r\to 0$,  exponentially      or powerlaw increasing  core factor and an, for $r\to\infty$,                  exponentially or powerlike decreasing tail factor. Concerning the subscripts,
we use an obvious notation when writing
  $$ \eqalign { U_{\alpha \beta}(r) & = \exp( {\alpha \rho_1\over r}- 
{\beta r\over \rho _2}),\cr  U_{a \beta}(r) & =(1+ {r_1\over r})^a 
 \exp( {-\beta r \over \rho_2}), \cr U_{\alpha b}(r)                           & =\exp({\alpha \rho_1 \over r}) ({r_2 \over r_2+r})^b, \cr
U_{ab}(r) & = (1+{r_1\over r})^a ({ r_2\over r_2+r})^b, \cr
[0\leq r, \ \ \  0< \alpha,\beta,& r_1,r_2,\rho_1,\rho_2; \ \  a>4; 
\ \  b>3].\cr}  \eqno(4.3) $$

In the last section, we established general criteria for the convergence
of the expansions (3.9a)-(3.9b). In the present section, fulfillment           of those requirements  will be checked  for increasing    nonlinear parameters.

          In the supersingularity limit $A \to \infty$,the asymptotical        form of the master equation (4.2) is explicitly solvable in cases             
of exponential tail for both types of core singularity. In fact, one concludes from the definitions (4.3) that
  $$  \eqalign{R_{\alpha \beta} & \to (\rho_1\rho_2)^{1\over 2} ({\alpha       
     \over\beta})^{1/2}, \ \ [\alpha\to \infty]; \cr                                 R_{a \beta} & \to (r_1\rho_2)^{1\over 2}({a \over \beta})^{1\over 2}, \ \ [a \to \infty]. \cr} \eqno(4.4) $$                                            In the powerlaw tail cases, in turn, only implicit expressions                    can be extracted from the formulas (4.2)-(4.3). Indeed, one gets
  $$ \eqalign {R_{\alpha b} & \to  r_2 \exp({\alpha \rho_1                     \over bR_{\alpha b}}), \ \ [\alpha \to \infty], \cr R_{a b}                    & \to r_2 \exp({ar_1 \over b R_{ab}}), \ \  [a\to \infty].} \eqno(4.5) $$      The matching  distance increases in all of our examples slower                 than  the respective variable parameter $A=\alpha \ \ {\rm or} \ \ a$.         Indeed,  one finds by the relationships (4.4)-(4.5) that
  $$ {R_{AB}\over A} \to 0, \ \ \ [A\to\infty;A=\alpha,a;B=\beta,b].           \eqno(4.6) $$

          The notation  $A\to \infty$ and $R_{AB}\to {\infty}$                 
compete in representing the supersingular limit by a single symbol.            We prefer the use of the latter alternative where possible. Accordingly,
we shall throughout  eliminate from the formulas the parameter $A$             in terms of $R_{AB}$.

          It is also obvious by analysis that the present                      scattering formalism should, in the supersingularity limit,  become,            mutatis mutandis, identical for exponentially and powerlaw                    increasing cores.  The equivalence is realized at the                          following  correspondence of potential  parameters:        
  $$ r_1a \to \rho_1 \alpha, \ \ [a,\alpha \to \infty,                         B=\beta,b]. \eqno(4.7) $$                                                      In  the limit considered, one has thus to treat, out of the four cases         in (4.3), only two essentially  different ones, namely ($A,\beta$) and        ($A, b)$. We now rewrite the wave number squares of Eqs.(2.8),                  in terms of the variable $t$ of the definition (2.4), for $R_{AB}\to\infty$.   In the cases ($A,\beta$), this transformation is, owing to the explicit        relationships (4.4), straightforward. Indeed, one obtains  by the                  definitions (4.3)  in  the exponential region 
  $$  \eqalign{ &K_{\epsilon A \beta}^2(t)  \to                                k^2 \{{g^2\over  k^2r_0^2} \exp[{\beta R_{A \beta} \over \rho_2}               ({1\over t}-t)]+ {\lambda_{\epsilon}^2 \over k^2R_{A \beta}^2t^2}-1  \},       \cr & \ \ [t<1,R_{A \beta}\to \infty, A=\alpha,a].\cr} \eqno(4.8) $$           A little bit more complicated is the incorporation of formula (4.5)            into  Eq.(2.8) in power tail cases for which one gets                          still in the  region $\epsilon$
  $$ \eqalign{ & K_{\epsilon A b}^2(t)  \to  k^2 \{[{k^2r_0^2                  \over g^2} ({R_{Ab} \over r_2})^b]^{{1\over t}-1} {1\over t^b} -1+                      {\lambda _\epsilon^2\over k^2 R_{Ab}^2 t^2} \},\cr                          &  \ \ [t<1,R_{Ab}\to \infty, A=\alpha,a].\cr} \eqno(4.9) $$ 
In the trigonometric region,  distinction should be made between               S-wave and  higher partial waves. Indeed, one extracts                         from  the definitions (2.8) and (4.3) for the potential                        classes $A=\alpha$ and $a$  equally,  that       
  $$ K_{\tau A \beta}^2(t) \to  k^2 \{ 1-{g^2 \over k^2r_0^2}                  \exp[-{\beta R_{A  \beta} \over \rho_2}(t-{1\over t})]\}, \ \                         [t>1,l=0,R_{A \beta} \to \infty], \eqno(4.10) $$
  $$ K_{\tau Ab}^2(t) \to k^2 \{1- [{g^2 \over k^2r_0^2}                       ({r_2 \over R_{Ab}})^b]^{1- {1\over t}}{1\over t^b} \}, \ \                                     [t>1,l=0,R_{Ab}\to \infty], \eqno(4.11) $$
  $$ K_{\tau AB}^2(t) \to k^2 \{1- {\lambda_\tau^2 \over k^
2R_{AB}^2t^2}\},    
\ \ [t>1,l>0,R_{AB} \to \infty]. \eqno(4.12) $$
Notice for  the highest values of the parameter $A$ the wave                     number function becomes independent of the potential:
  $$ K_{\tau AB}^2(t)\to k^2, \ \ [t>1,l\geq 0,R_{AB}\to\infty].               \eqno (4.13) $$

          The quantities $D^{(s)}_\gamma(t)$ of the definition (2.12 will      be  calculated below from the set of supersingularity expressions              (4.8)-(4.12). In the exponential region one finds for both cases               $A=\alpha,a$  in  the limit $R_{AB}\to\infty$
  $$ D_{\epsilon A \beta}^{(1)}(t) \to -{\beta R_{A \beta} \over \rho_2}       ({1\over t^2}+1), \ \                                                          D_{\epsilon A \beta}^{(2)}(t) \to ({\beta R_{A \beta}\over \rho_2})^2          ({1 \over t^2}+1)^2, \ \ [t>1,l\geq 0],\eqno(4.14) $$
  $$ D_{\epsilon A b}^{(1)}(t) \to -{b\over t^2}                               
\ln({R_{Ab}\over r_2}), \ \
D_{\epsilon A b}^{(2)}(t)\to {b^2\over t^4}[\ln ({R_{Ab}\over r_2})]^2,          \ \  [t>1,l\geq 0]. \eqno(4.15) $$
As regards the trigonometric region, the formulae are again sensitive          to the orbital angular momentum. Indeed,                                       for the potential classes $A=\alpha,a$ alike, one gets
  $$ D_{\tau A \beta}^{(1)}(t)\to {\beta R_{A\beta}\over \rho_2}               ({1\over t^2}+1), \ \  D_{\tau A\beta}^{(2)}(t) \to                            -({\beta R_{A \beta}\over \rho_2})^2
({1\over t^2}+1)^2, \ \ [t>1,l=0],\eqno(4.16) $$
  $$ D_{\tau Ab}^{(1)}(t)\to {b\over t^2} [\ln ({R_{Ab}\over r_2})]^2, \ \ 
D_{\tau Ab}^{(2)}(t)\to -{b^2\over t^4}[\ln({R_{Ab}\over r_2})]^2, \ \ [t>1,   l=0],
\eqno(4.17) $$
  $$ D_{\tau AB}^{(1)}(t)\to {2\lambda_{\tau}^2\over k^2 R_{AB}^2t^3}, \ \         D_{\tau AB}^{(2)}(t)\to -{6\lambda_{\tau}^2\over R_{AB}^2t^4}, \ \      
[t>1,l>0].\eqno(4.18) $$

          The residual potentials (2.23)-(2.24) are quadratic and linear       
expressions of the quantities (4.14)-(4.18). They read  in                     the exponential region, for exponential and powerlaw cores in  like manner,      $$ \Delta_{\epsilon A \beta}(t)\to-{1\over 16}                                ({\beta R_{A\beta}\over \rho_2})^2({1\over t^2}+1)^2,                          \ \ [t<1,R_{A \beta} \to \infty], \eqno(4.19) $$
  $$ \Delta_{\epsilon A b}(t)\to -{1\over 16} [\ln({R_{Ab}\over r_2})]^2     
{b^2\over t^4}, \ \ [t<1,R_{Ab} \to \infty]. \eqno(4.20) $$
The analogous formulae of the trigonometric region are sensitive to the         orbital angular momentum as
  $$ \Delta_{\tau A\beta}(t)\to -{9\over 16}                                   
({\beta R_{A\beta} \over \rho_2})^2
({1\over t^2}+1)^2, \ \ [t>1,l=0,R_{A \beta}\to \infty], \eqno(4.21) $$
  $$ \Delta_{\tau Ab}(t) \to -{9 \over 16} [\ln({R_{Ab} \over r_2})]^2
{b^2\over t^4}, \ \ [t>1,l=0, R_{Ab} \to \infty], \eqno(4.22) $$
  $$ \Delta_{\tau AB}(t)\to {3\lambda_{\tau}^2\over 2k^2R_{AB}^2t^4},           \ \ \ [t>1,l>0,R_{AB}\to \infty,B= \beta,b]. \eqno(4.23) $$

          The next step towards checking expansions (3.9) for the              realization of convergence criteria is calculation of the quantities           $p_{\gamma}(t)$  introduced by Eq.(3.12). The asymptotical  expressions        (4.8)-(4.13)  combine with the ones of (4.19)-(4.23) to yield,  first          for the exponential region,                                                       $$ p_{\epsilon A\beta}(t)\to -{1\over 16} {\beta^2R_{A\beta}r_0             
\over g\rho_2 ^2}({1\over t^2}+1)^2\exp[-{\beta R _{A\beta}\over                              \rho_2}({1\over t}-t)],
 \ \ [t<1,R_{A \beta} \to \infty], \eqno(4.24) $$
  $$ p_{\epsilon Ab}(t)\to -{1\over 16}[\ln ({R_{Ab}\over r_2})]^2                {b^2 \over kR_{Ab}}t^{{b\over 2}-2} [{g^2\over k^2r_0^2}                    ({r_2\over R_{Ab}})^b]^{{1\over t}-1}, \ \ [ t<1,R_{Ab} \to \infty],           \eqno(4.25) $$
as well as in the trigonometric region                                          $$ p_{\tau A\beta}(t)\to -{9\over 16}({\beta R_{A \beta} \over \rho_2})^2    {1\over kR_{A\beta}} ({1\over t^2}+1)^2, \ \                                   
[t>1,l=0,R_{ A \beta} \to \infty],\eqno(4.26) $$
  $$ p_{\tau Ab}(t)\to -{9\over 16}[\ln({R_{Ab}\over r_2})]^2{b^2\over t^4}
{1\over kR_{Ab}}, \ \ [t>1,l=0,R_{Ab} \to \infty],\eqno(4.27) $$
  $$ p_{\tau AB}(t)\to {3\lambda_{\tau}^2\over 2k^2R_{AB}^2t^4}
{1\over kR_{AB}}, \ \ [t>1,l>0,R_{AB} \to \infty].\eqno(4.28) $$
The supersingularity  forms of  $P_\epsilon(0,t)$ and $P_\tau(1,t)$            are obtained  by first integrating  the expressions (4.24)-(4.28)               over the relevant  intervals and subsequently going over                      
 to the limit $R_{AB}\to\infty$.
 
Within the exponential region, the limits $t\to 0$ and $R_{AB}\to\infty$      
mutually strengthen the rate of vanishing. The factor $({r_2\over R_{Ab}})      ^{1\over t}$ in the expression (4.25) vanishes at
$R_{Ab}>r_2$ for $t\to 0$ . The functions
$p_{\epsilon AB}(t)$ are  integrable  near the point $t=0$  even at             finite values of  $R_{AB}$. Moreover, its  integral vanishes                   in the limit $R_{AB}\to \infty$. Thus
  $$ P_{\epsilon AB}(0,t) \to 0, \ \ \ [t<1; A=\alpha, a;R_{AB}\to\infty].
\eqno(4.29) $$
By virtue of the inequality (3.17), the condition  for the   absolute          convergence of the series  (3.9a) is  thus  fulfilled  in the
supersingularity  limit along the exponential region for each of the           potentials $U_{AB}(r)$ of the set (4.3).              

As regards the trigonometric region, there is a competition between            the potential and the centrifugal term and this is the point that              
governs convergence.In the absence of the latter, one extracts                 from Eq.(4.26) for both cases $A=\alpha,a$  that
  $$ P_{\tau A\beta}(1,t)\to \infty,\ \ [t>1;l=0;R_{A\beta}\to\infty].          \eqno(4.30) $$
The convergence of the series (3.9b) is thereby frustrated for the S-wave      whenever the physical  potential decreases exponentially. Not so for           the higher partial waves or the cases of  powerlaw tails. Equations            (4.27)-(4.28) imply namely that
  $$ P_{\tau Ab}(1,t) \to 0,\ \ \ [t>1;l=0;R_{Ab}\to\infty],\eqno(4.31) $$
  $$ P_{\tau AB}(1,t) \to 0,\ \ \ [t>1;l>0;R_{AB}\to\infty].\eqno(4.32) $$
The asymptotical relationships (4.29), (4.31) and (4.32) ensure, by the        the inequalities  (3.17) and  (3.22), for the  respective potential            classes and  partial waves, fulfillment of the convergence conditions         (3.23), simultaneously at fixed and increasing values of the nonlinear         parameter $A$, involved in the scattering potential $U_{AB}(r)$.               One can thus write in the limit $R\to\infty$ that
  $$  u^+(t)\to v^+_\epsilon(t), \ \ [t<1; B=\beta,b;l\geq 0], \eqno(4.33) $$
  $$  u^+(t)\to v^+_\tau(t), \ \ [t>1;B=b;l=0],\eqno(4.34) $$
  $$  u^+(t)\to v^+_\tau(t), \ \ [t>1;B=b,\beta;l>0]. \eqno(4.35) $$           \bigskip
\centerline {\bf  V. Asymptotical exactness}                                              In the last section we studied the question whether a               semiclassical expansion that is convergent at a fixed set of dynamical         parameters  preserves this property invariably in the supersingular limit.     A further point we are left with to clear is  how the structure of the         series changes upon going with the nonlinear parameter to infinity.            A characteristic quantity of the argument for finding the answer               is the ratio of two neighboring general terms.  Recall first the               definition (3.11) of $q_{\gamma s}^+(t)$ along with  the relevant              
recursion relationships (3.14) and (3.18). For simplicity, in the              present section we are going to suppress in the formulas the matching distance $R$. Remember it is a functional of the scattering potential                   $g^2U_A(Rt)$ and is, in fact,
invisibly present in each quantity and expression below.

        In the exponential region, the exact formula (3.14) reduces            in  the supersingular limit, owing to the definition (2.17), to                  $$q^+_{\epsilon n}(t) \to  \int_0^t{\rm d}t'p_\epsilon(t')                 
q_{\epsilon n-1}^+(t'), \ \ \ [t<1,R\to\infty].\eqno(5.1) $$
Iteration yields then  in the limit considered a  familiar expression
  $$  q^+_{\epsilon n}(t)\to {1\over n!}[P_\epsilon(0,t)]^n, \ \               [t<1,R\to\infty]. \eqno(5.2). $$
The scattering wave function reads thus by the definitions (3.9a)               and (3.13) in the limit discussed      
  $$ v_\epsilon^+(t)\to  w_{\epsilon 0}^+(t) \exp {P_\epsilon (0,t)}, \ \     
[t<1,R\to\infty]. \eqno(5.3) $$                                                For treating the trigonometric region , it is useful to  introduce higher      order trigonometric 'coefficients', which  are, as a matter of fact,           no more constant. The definition is meant to hold for both fixed or            variable dynamical parameters and reads
  $$ w_{\tau m}^+(t)= \eta_\tau (t) [C_m^+(t)\cos \omega_\tau(1,t)
  +S_m^+(t) \sin \omega_\tau(1,t)], \ \ [t>1,m=0,1..].\eqno(5.4) $$           
Recognize the identity (5.4) reproduces at $m=0$ the definition (2.15),
on account of which  one finds that $C_0^+=C^+$ and $S_0^+=S^+$ . 
Combination of this identity with the recursion formula (3.18)                 furnishes in our limit, after separating the sine- and cosine-                 contributions, the following  system of equations        
  $$  C_m^+(t)\to \int_1^t{\rm d}t'p_\tau(t')\sin[kR(t'-1)]
w_{\tau m-1}^+(t'), \ \ [t>1,R\to\infty],\eqno (5.5) $$ 
  $$  S_m^+(t) \to\int_1^t{\rm d}t'p_\tau(t')\cos[kR(t'-1)]
w_{\tau m-1}^+(t'), \ \ [t>1,R\to\infty].\eqno(5.6) $$
Insertion of the definition (5.4) into the right hand  sides                  
of the last  two formulas  yields by analysis, owing  to the                   infinitely rapid oscillations  in the integrands, a  pair of                   coupled systems  of integral equations such as
  $$ C_m^+(t)\to {1\over 2}\int _1^t {\rm d}t'p_\tau(t')S_{m-1}^+(t'), \ \     [t>1,R\to\infty], \eqno(5.7) $$
  $$ S_m^+(t)\to -{1\over 2}\int_1^t{\rm d}t'p_\tau(t')                          C_{m-1}^+(t'), \ \ [t>1,R\to\infty]. \eqno(5.8) $$
At the end of the iteration, at $m=0$, one encounters  the  constant             coefficients $C_0^+,S_0^+$ . The system  of integral equations                 becomes thereby explicitly  solvable. The solution reads, in terms             of the notation 
  $$  T_m(t)\equiv {1\over m!}[{1\over 2}P_\tau(1,t)]^m, \ \ [t>1,m=0,1,2,..],
\eqno(5.9) $$                                                                  as follows:
  $$ C_{4\mu}^+(t)\to +T_{4\mu}(t)C_0^+, \ \ \ \ \ S_{4\mu}^+(t)\to                +T_{4\mu}(t)S_0^+,\eqno(5.10) $$
  $$ C_{4\mu+1}^+(t)\to +T_{4\mu+1}(t)S_0^+, \ \ \ \ \                             S_{4\mu+1}^+(t)\to -T_{4\mu+1}(t)C_0^+,\eqno(5.11) $$
  $$ C_{4\mu+2}^+(t)\to -T_{4\mu+2}(t)C_0^+, \ \ \ \ \   S_{4\mu+2}^+(t)          \to -T_{4\mu+2}(t)S_0^+, \eqno(5.12) $$
  $$ C_{4\mu+3}^+(t)\to -T_{4\mu+3}(t)S_0^+,  \ \  \ \ \  S_{4\mu+3}^+(t)           \to +T_{4\mu+3}(t)C_0^+,\eqno(5.13) $$
where  $\mu=0,1,2,..$.                                                         
Incorporation of the formulas (5.10)-(5.13) into the definition (3.9b)         
furnishes
  $$  \eqalign {v_\tau^+(t)\to \sum_{\mu=0}^\infty
\{ & [T_{4\mu}(t)-T_{4\mu+2}(t)] w_{\tau 0}^+(t)+ \cr
   & [T_{4\mu+1}(t)-T_{4\mu+3}(t)] w_{\tau 0}^-(t)\},\ \ [t>1,R\to\infty],
\cr}\eqno(5.14) $$
where we used the ad hoc yet not inconsistent notation                        
  $$ w_{\tau 0}^-(t) =\eta_\tau(t) \{S_0^+ \cos[kR(t-1)]-
C_0^+ \sin [kR(t-1)]\} , \ \ [t>1], \eqno(5.15) $$

Observe there is only a single quadrature involved in the asymptotical          formula  (5.14), namely the one implicitly contained in the definition (5.9).   Knowledge of the higher order $t$-dependent coefficients thus rests upon     
the knowledge of the $R\to\infty$  form of the zero order constants
$C^+_0$ and $S^+_0$. These constants are fixed by the claim for                the  smoothness of the exact but supersingularity  solution                    
   at $t=1$ as follows.

At the matching point itself, approaching it from either of the regions         $\epsilon$ or $\tau$, one obtains the respective  amplitude functions,         see Eq.(2.16), along with the relevant derivatives  as follows:
  $$ [\eta_\gamma(t)]_{t=1}\to 2^{{3\over 4}} (kR)^{{1\over 2}},             \ \ \ [\gamma=\epsilon,\tau; \ \ R\to\infty] ,\eqno(5.16) $$
  $$ [{{\rm d} \eta_\gamma(t) \over {\rm d} t}]_{t=1} \to \pm 2^{{7\over 4}}  
(kR)^{{1\over 2}} [2\lambda_\gamma^2 - g^2R^3U'(R)], \ \ [\gamma=(^\epsilon
_ \tau); \ \ R\to\infty]. \eqno(5.17) $$
The second term in the brackets vanishes in this limit by the restrictions 
imposed upon the potentials in virtue of the asymptotical                      relationships (2.1).  

     As to the  exponential region, the supersingularity formula (5.3)        
can be identically recast by means of the definition (2.14) as

   $$  v_\epsilon^+(t) \to \eta_\epsilon(t) \exp{[\omega_\epsilon(1,t)
+P_\epsilon(0,t)]}, \ \ [t<1,R\to\infty]. \eqno(5.18) $$
Approaching  the matching point from this region , one concludes by             Eqs. (5.16)-(5.18) that 
  $$  [v_\epsilon^+(t)]_{t=1}\to 2^{3\over 4}(kR)^{1\over 2}                \exp{P_\epsilon (0,1)}, \ \ [R\to\infty], \eqno(5.19) $$
  $$  [{{\rm d} v_\epsilon^+(t)\over {\rm d}t}]_{t=1} 
\to 2^{3\over 4} (kR)^{1\over 2} \exp P_\epsilon(0,1)                          \{4\lambda_\epsilon^2+2^{-{3\over 2}}+[p_\epsilon(t)]_{t=1}\},
 \ \ [R\to\infty].\eqno(5.20) $$
Recall now the expressions (4.24)-(4.25) and extract  from them
  $$ [p_{\epsilon }(t)]_{t=1}={\cal  O} \{{1\over kR}\}\to 0, \ \      
[B=b;R\to \infty], \eqno(5.21) $$
  $$ [p_{\epsilon}(t)]_{t=1}= {\cal O}\{{R\over \rho_2} \}
\to\infty, \ \ [B=\beta;R\to \infty] . \eqno(5.22) $$
Therefore, we  have to restrict further discussion to scattering by            potentials of the class $U_{Ab}(r), [A=\alpha, a]$ of Eqs. (4.3).
As regards the trigonometric region,  the definition (2.15) yields
  $$  [w^+_{\tau 0}]_{t=1}\to 2^{3\over 4} (kR)^{1\over 2} C^+_{0},              \ \ \ [B=b,R\to\infty], \eqno(5.23) $$
  $$  [{{\rm d}w_{\tau 0}^+(t)\over {\rm d} t}]_{t=1}\to (kR)^{1\over 2}       [-2^{11\over 4} \lambda_\tau^2 C^+_{0}+S^+_{0}],  \ \ \  [B=b,R\to\infty].     \eqno(5.24) $$ 
Smooth matching is realized by equating Eqs.(5.19) and (5.23)                  as well as (5.20) and (5.24), respectively. In doing so, one obtains
  $$  C^+_{0}\to \exp{P_\epsilon(0,1)}, \ \ \ 
      S^+_{0}\to \exp{P_\epsilon(0,1)}[2^{{15\over 4}}                         \lambda^2+2^{-{3\over 4}}], \ \ [B=b,R\to\infty]. \eqno(5.25) $$
As to the region $\epsilon$, Eqs. (4.29), (4.33) and (5.3) , as well as
concerning the region $\tau$,  Eqs. (4.31), (4.32), (5.9) and (5.14),
combine to the statement that , for potentials  $U_{AB}(r)$                    with powerlaw tails, [B=b],  the pair of zero order terms                      in the semiclassical expansions of the  respective regions  themselves 
recover in the supersingular limit  the exact QM solution. In particular,
  $$  u^+(t)\to w^+_{\epsilon 0}(t), \ \ \ [t<1;B=\beta,b;R\to\infty],         \eqno(5.26) $$
  $$  u^+(t)\to w_{\tau 0}^+(t), \ \ \ [t>1;B=b;  l \geq 0];R\to\infty].
\eqno(5.27) $$
 In cases of  exponential potential tails, the S-partial waves should           be excluded from the present approach.             
\bigskip
\centerline {\bf VI. Discussion}

         Usefulness of a modified semiclassical approach in treating           $singular$  $scatter\-ing$  has been checked in three steps. First,            
the conditions  were reconsidered at which a smooth WKB method,                proposed recently,produces convergent expansion of the wave function           at fixed set of the  potential  parameters. An inherent  new point is          that, instead of a single one, a $pair$ of $integral$ $equations$              should be set up, one for each of  the exponential and the                     trigonometric regions. A  regular solution working in the exponential region   selects by virtue of  the  smoothness postulate a particular one               out of the solutions prepared  for the trigonometric region.                   Another new item is the extension  
of the argument over potentials involving  $variable$   $nonlinear$            $parameters$ through  the variation of  which one  may increase                 the core  singularity  to  asymptotically high levels. Analyzed are           four classes of interactions  each of which is a product of                    a core factor implying  exponential  or powerlaw singularity  and of           a tail factor that decays either exponentially or powerlike. Independently   
of the stage of the singularity, the powerlike decaying potentials             invariably  develop absolute convergent expansions, along both regions,        also  in  the supersingularity limit. In  scattering by exponentially           decaying potentials, the criteria of convergence seemingly fail               to work within our argument of treating supersingularity limit.                A further  new feature of the present approach  is a  discussion  of the        $quality$ of $convergence$. The conditions are found  which  the shape of     the potential has to show up  so that the infinite series shrink, in the          supersingularity  limit, to a single term. It is, perhaps,                  
 worth mentioning that,  by  varying the length of the cut-off series           in the region  beyond the matching point, one can obtain, at the expense       of a single quadrature, a solution  that  becomes correct to any               prescribed order in the reciprocal supersingularity parameter                  at its asymptotically large values.

Recall the Born series furnishes the physical scattering wave function for     nonsingular potentials at fixed linear and nonlinear dynamical                 parameters, exclusively. The proposed smooth WKB approach should work          for singular potentials at fixed and asymptotical values of linear and         nonlinear parameters involved in the 
Schr\"o\-din\-ger equation.                                                                                                                                   
\bigskip
\parindent 0pt

{\bf {Acknowledgement}}
Many thanks are due to Dr. G.Bencze for useful critical remarks. The
author is grateful to Dr. G.Kluge and Dr. I. Racz for very valuable
discussions. The work was partly supported by the Hungarian NSF under
Grant No. OTKA 00157.

{\bf {References }}
\bigskip
\parindent 0pt
\baselineskip=9.5 pt
\parskip=2 pt 

\item{$^1$} F. Calogero: Variable Phase Approach to Potential Scattering,                  (Academic Press, {\hskip 3cm} {New York, 1967)} 
\item{$^2$} F. Calogero: Phys. Rev. {\bf B 135}, 693 (1964)
\item{$^3$} G. Esposito: J.Phys {\bf A 31}, 9493 (1998)
\item{$^4$} N. Froeman and K.-F. Thylwe: J. Math. Phys  {\bf 20}, 1716 (1979)
\item{$^5$} R.G. Newton: Scattering Theory of Waves and Particles,                          (Springer Verlag, 1981) 
\item{$^6$} T. Dolinszky: Physics Letters {\bf A 132}, 69 (1988)
\item{$^7$} T. Dolinszky: J.Math.Phys. {\bf 36}, 1621 (1995)
\item{$^8$} T. Dolinszky: J.Math.Phys. {\bf 38}, 16 (1997).

\vfill
\end